\documentclass{article} 
\usepackage{iclr2023_conference,times}

\usepackage{amsmath,amsfonts,bm}

\def\eqref#1{equation~\ref{#1}}

\def\floor#1{\lfloor #1 \rfloor}
\def\1{\bm{1}}

\DeclareMathAlphabet{\mathsfit}{\encodingdefault}{\sfdefault}{m}{sl}
\SetMathAlphabet{\mathsfit}{bold}{\encodingdefault}{\sfdefault}{bx}{n}

\usepackage{hyperref}
\usepackage{url}
\usepackage{multirow}
\usepackage{tabularx}
\usepackage{threeparttable}
\usepackage{xspace}
\usepackage{booktabs}
\usepackage[thinc]{esdiff}
\usepackage{graphicx}
\usepackage{comment}
\newcommand{\pd}[2]{\frac{\partial{#1}}{\partial{#2}}}
\newcommand{\pdd}[2]{\frac{\partial^2{#1}}{\partial{#2}^2}}
\newcommand{\sys}{\texttt{NetFlick}\xspace}

\newcommand{\norm}[1]{\displaystyle \left\| #1 \right\|}

\title{\sys: Adversarial Flickering Attacks on Deep Learning based Video Compression}

\author{Jung-Woo Chang$^1$, Nojan Sheybani$^1$, Shehzeen Hussain$^1$, \textbf{Mojan Javeheripi}$^1$, \\
\textbf{Seira Hidano}$^2$, \textbf{Farinaz Koushanfar}$^1$ \\
$^1$University of California, San Diego \\
\texttt{\{juc023, nsheyban, ssh028, mojan, farinaz\}@ucsd.edu} \\
$^2$KDDI Research, Inc. \\
\texttt{\{se-hidano\}@kddi.com} \\
}

\iclrfinalcopy 
\begin{document}

\maketitle
\vspace{-5mm}
\begin{abstract}

Video compression plays a significant role in IoT devices for the efficient transport of visual data while satisfying all underlying bandwidth constraints. Deep learning-based video compression methods are rapidly replacing traditional algorithms and providing state-of-the-art results on edge devices. However, recently developed adversarial attacks demonstrate that digitally crafted perturbations can break the Rate-Distortion relationship of video compression. In this work, we present a real-world LED attack to target video compression frameworks. Our physically realizable attack, dubbed \sys, can degrade the spatio-temporal correlation between successive frames by injecting flickering temporal perturbations. In addition, we propose universal perturbations that can downgrade performance of incoming video without prior knowledge of the contents. Experimental results demonstrate that \sys can successfully deteriorate the performance of video compression frameworks in both digital- and physical-settings and can be further extended to attack downstream video classification networks.
\end{abstract}

\vspace{-5mm}
\section{Introduction}

Deep Neural Networks (DNNs) are utilized for a myriad of media-related tasks, such as video and image classification~\cite{carreira2017quo, feichtenhofer2019slowfast} and audio transcription~\cite{gupta2014vector}. Over the past years, DNNs have evolved in performance, but several works have shown their susceptibility to adversarial perturbations. Initial works~\cite{szegedy2013intriguing, carlini2017towards} in this field have shown that image classification systems can be attacked with carefully crafted and imperceptible adversarial additions to the inputs. Furthermore, 
physical adversarial examples such as adversarial patches~\cite{eykholt2018robust,chen2019shapeshifter}, have been utilized to disrupt real-world image classification models deployed in autonomous cars. 


For the success of media-related tasks, video compression plays a significant role in minimizing redundancy for video content delivery services, e.g., video surveillance~\cite{wang2018background}, AR/VR~\cite{jang2019video}, remote surgery~\cite{hassan2019high}, etc. Video compression typically follows \textit{R-D} optimization that minimizes the distortion (\textit{D}) at a given bit rate ($\textit{R}<\textit{R}_{t}$), where $\textit{R}_{t}$ is the available bit rate budget. With this optimal trade-off between video quality and bit rate, video streaming and classification systems can handle four-step workflows: (1) Input data from a front-end video sources (camera) (2) Video encoder to compress data into a bitstream for communication to the final destination (3) Video decoder to decompress bitstream back into video format (4) Downstream services (streaming and classification). Recently, DNN-based video compression ~\cite{Lu_2019_CVPR} has been explored by Moving Picture Experts Group (MPEG) for adoption in the next-generation of video delivery systems~\cite{Mpeg} due to its higher performance than conventional codecs~\cite{sullivan2012overview}. However, \cite{chang2022rovisq} have shown that DNN-based video compression is vulnerable to imperceptible adversarial perturbations which when added to digital input video frames, can manipulate the \textit{R-D} relationship. When considering video compression, studies that realize attacks in the physical world have not yet been investigated.

In this work, we present a physical adversarial attack- \sys as a new threat to video compression systems in the real-world. 
We find that the carefully crafted flickering effect of \sys using WiFi-controlled RGB LED light bulb, can severely threaten video compression efficiency. 
In summary, our contributions are:
\vspace{-2mm}
\begin{itemize}
    \item We propose \sys, a novel physical attack to video compression systems, relying on adversarial flickering perturbations via a smart RGB LED light bulb.
    \item We evaluate the \sys attack scheme on several video compression and downstream video classification frameworks to highlight its effectiveness.
    \item We present online and offline attack schemes that both heavily degrade compression and classification, alongside a  physical framework to realize \sys in the real world.
\end{itemize}






\section{Preliminaries}
\subsection{Deep Learning-based Video Compression}
Video compression techniques focus on predicting the temporal relationships between frames in order to minimize the amount of data that needs to be transmitted. The encoder generates a stream of bits that conform to a specific channel standard, which is then sent to the decoder. The decoder then uses this stream of bits to reconstruct the compressed video.
Over the past few decades, many video compression standards have been introduced, such as VP8~\cite{bankoski2011technical}, H.264~\cite{wiegand2003overview} and H.265~\cite{sullivan2012overview}. Traditionally, techniques for reducing the spatial and temporal redundancies in video sequences relied on manually crafted algorithms. Recently, the use of DNNs in video compression frameworks, such as those proposed by \cite{Lu_2019_CVPR, Hu_2021_CVPR, Yang_2020_CVPR}, has gained significant attention due to its superior performance. This relies on three main design components: (1) motion estimation network for predicting the temporal motion, (2) motion compensation network to generate the predicted frame, and (3) auto-encoder based network for compressing the motion and residual data into bitstream.

\subsection{Adversarial Attacks}
Several existing studies~\cite{szegedy2013intriguing, carlini2017towards,moosavi2017universal, shamsabadi2020colorfool, qiu2020semanticadv, Hussain_2021_WACV,pony2021over} have demonstrated the vulnerability of DL based-image and video classification models to adversarial attacks. These attacks can be either digital~\cite{szegedy2013intriguing, moosavi2017universal} or physical~\cite{chen2019shapeshifter, lovisotto2021slap} depending on their feasibility in the real world. To be specific, \cite{chen2019shapeshifter} printed out perturbed traffic signs and pasted them onto victim traffic sign boards to conduct physical attacks on traffic sign recognition systems. \cite{lovisotto2021slap} projected adversarially crafted physical perturbations onto real-world objects using a projector.
Recently, \cite{chang2022rovisq} demonstrated the first adversarial attacks on video compression to manipulate the \textit{R-D} relationship in a digital setting. 
However, studies on the effectiveness of physical adversarial attacks to video compression have not yet been conducted.




\section{Methodology}

\subsection{Victim Models}
To demonstrate the effectiveness of the proposed \sys attack on video compression, we choose
state-of-the-art models that employ temporal prediction to minimize the difference between a previously decoded video frame and the current video frame. 
Let $X=[x_{1}, \dots, x_{T}] \in \mathbb{R}^{T \times W \times H \times C}$ denote a video clip containing $T$ consecutive frames, where $x_{t}$ is the frame at a time step $t$ which has $H$ rows, $W$ columns, and $C$ color channels. 
The goal of a victim DL-based video compression
model
is to efficiently remove the spatio-temporal information within the $g$-th group of pictures (GOP), where $0\leq g \leq \lfloor \frac{T}{G} \rfloor$ and $G$ is the total number of frames in the GOP. Each frame is efficiently represented by coding a difference from a reference frame, rather than repeatedly coding each individual frame.

Accordingly, we formulate the video encoder as a function $\mathbb{E}(x_{t}, \mathcal{P}_{t}, \lambda) = b_{t}$ that compresses the current frame $x_{t}$ into a bitstream $b_{t}$ according to a compression rate $\lambda$ and reference frames $\mathcal{P}_{t}$ which are previously compressed. Here $\mathcal{P}_{t}=[y_{G \cdot g + 1}, \dots,  y_{G \cdot g + G}]$ where $y_{(\cdot)}$ is the output of the decoder at a certain time step. Note that $\mathcal{P}_{1}=\emptyset$ since the first frame, dubbed the I-frame, is coded independently using DL-based image compression. Video decoder can be formulated as a function $\mathbb{D}(b_{t}, \mathcal{P}_{t}, \lambda) = y_{t}$ that reconstructs the decoded frame $y_{t}$ from the bitstream $b_{t}$ using the reference frames $\mathcal{P}_{t}$ at a given compression rate $\lambda$. Optionally, the downstream video classification service receives as input the decoded frames from the video compression pipeline.


\begin{figure}[h]
\centering
    \begin{tabular}{@{}cc@{}}
    \includegraphics[width=0.5\linewidth]{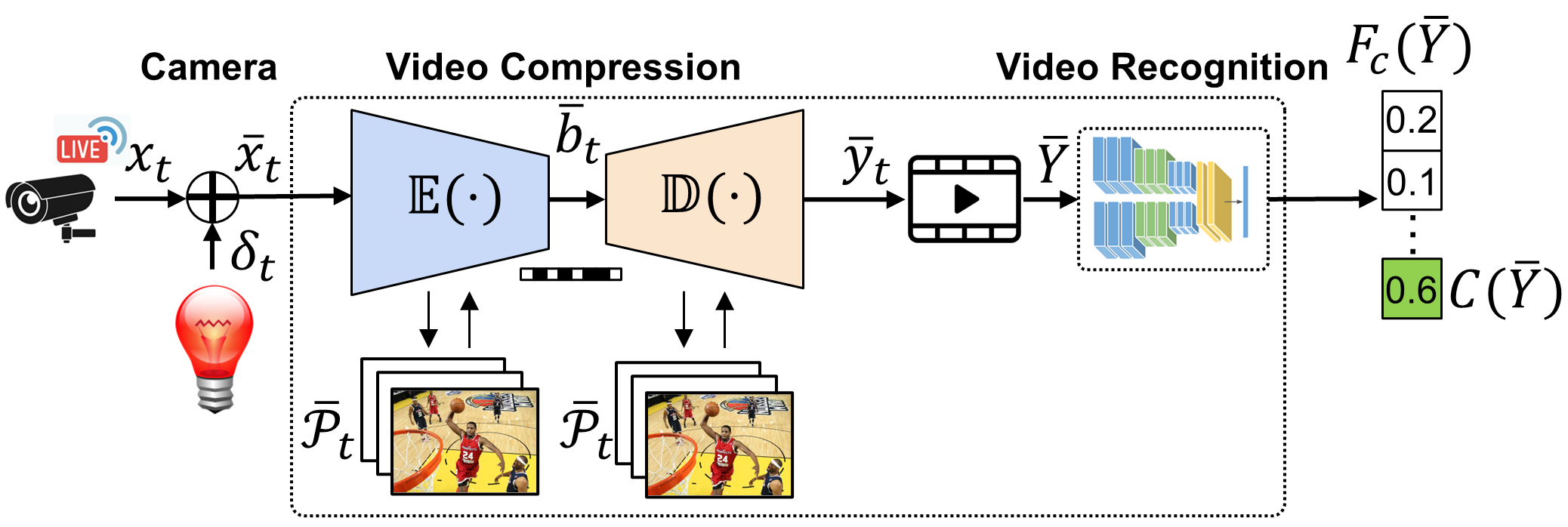} &
    \includegraphics[width=0.4\linewidth]{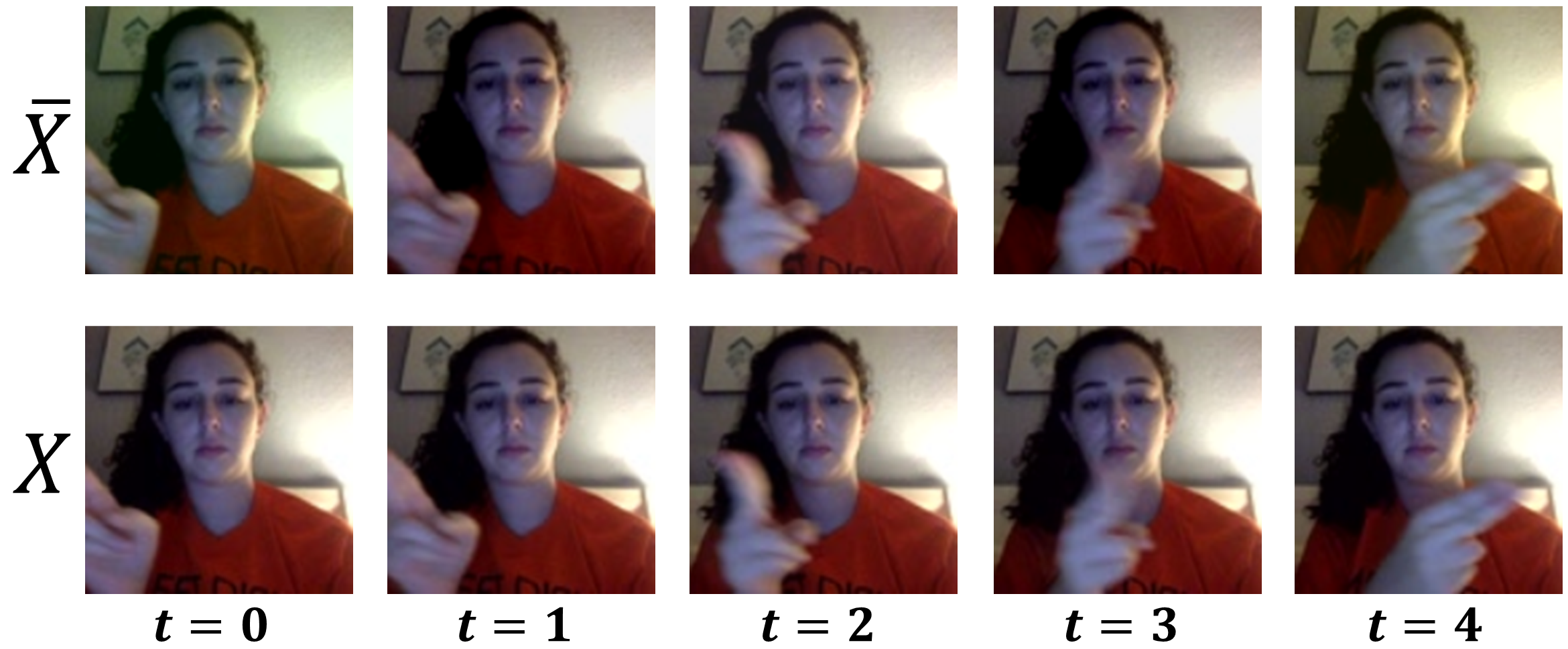} \\
    (a) &
    (b) \\
    \end{tabular}  
  \caption{(a) Overview of system components for the proposed \sys attack on video compression and recognition. (b) Visual comparison between adversarial video ($\bar{X}$) and clean video($X$).}
  \label{fig:attack}
  \vspace{-0.5cm}
\end{figure}

\subsection{Threat Model}

\textbf{Attack Goal: } 
IoT devices such as surveillance cameras are equipped with a microprocessor for video compression. 
\sys attack aims to create physical adversarial perturbations using LED bulb flicker to target neural video compression frameworks that are optionally followed by video classification models. 
While prior work~\cite{pony2021over} has shown adversarially crafted flickering light to be an effective attack on video classification, they do not consider any video compression framework in their threat model.
We extend this work by formulating the first physically realizable attack to compromise both the video compression and video classification performance using the attack objective described in Section~\ref{sec:offlineattack} and Section~\ref{sec:onlineattack}.


\textbf{Attack Scenarios: } 
We consider two attack scenarios- online and offline. In an offline attack setting, we assume that the adversary 
can arbitrarily inject perturbations into the target frame by an RGB LED bulb. 
In an online attack setting, the adversary performs untargeted attack using an RGB LED bulb to negatively impact the video compression performance of an IoT device located in the same room.
 Specifically, the attacker installs a WiFi-controlled RGB LED bulb in a room to interfere with a camera of a victim device situated in the same room, by projecting adversarially crafted light onto it when a person streams. 
 An overview of the \sys framework to video compression modules is presented in Figure~\ref{fig:attack}(a), along with  corresponding inputs, outputs and the perturbation. 
 
\textbf{Adversary's Capabilities: }For offline attacks, we assume a white-box attack scenario where the architecture and weights of the video compression model are known to adversary. This is a realistic assumption since video compression models are standardized by various organizations, e.g., ISO/IEC and MPEG, and are open-source to the community. However, in reality, it is difficult for attackers to fully grasp the structure and all parameters of the video compression model and classification model. Thus, we consider a black-box attack scenario for online attacks, i.e. the adversary does not have knowledge about the architecture or parameters of the victim DNN models. In addition, we assume that the adversary has access to a public dataset to train the online attack.

\subsection{Offline Attacks}
\label{sec:offlineattack}

\textbf{Adversarial loss.} Let $\Delta = [\delta_{1}, \dots,\delta_{T}]$ $\in \mathbb{R}^{T\times 1 \times 1 \times C}$ denote flickering perturbations for a given video $X$. $\Delta$ is designed to be spatial-constant on the color channels, so $\delta$ can be represented by three scalars.
We denote the resulting adversarial video by $\bar{X}=[\bar{x}_{1},\dots, \bar{x}_{T}] \in \mathbb{R}^{T\times W \times H \times C}$ which consists of adversarial frames $\bar{x}_{t}=x_{t}+\delta_{t}$. Upon receiving the adversarial input $\bar{x}_{t}$, the video encoder outputs an adversarial bitstream $\bar{b}_{t}$ using perturbed reference frames stored in the the buffer $\bar{\mathcal{P}}_{t}= [\bar{y}_{G \cdot g+1}, \dots, \bar{y}_{G \cdot g+G)}]$. Finally, the video decoder restores the adversarial decoded frame $\bar{y}_{t}$ from the perturbed bitstream $\bar{b}_{t}$. The objective of our attack is to find $\Delta$ that can simultaneously target \textit{R} and \textit{D} to increase the bit rate and video distortion as follows: 
\vspace{-2mm}
\begin{equation} \label{eq:eq1}
\begin{aligned}
\min_{\Delta_{g}} \ \mathcal{L}_{comp}(X, \Delta_{g}, \lambda, g), \quad\quad
\mathcal{L}_{comp}(X, \Delta_{g}, \lambda, g) = -\sum^{G \cdot g+G}_{t = G \cdot g + 1} (R(\bar{b}_{t}) + \lambda \cdot D(x_{t}, \bar{y}_{t})), \\
\end{aligned}
\vspace{-1.5mm}
\end{equation}

where $\Delta_{g}$ is the perturbation for the $g$-th GOP.

After decoding the adversarial video, the back-end user may use a video classification as the target downstream task. Let $F(\bar{Y})$ denote a discriminant function that outputs a probability distribution over a set $K$. $F_{c}(\bar{Y})$ is the probability that $\bar{Y}$ belongs to a specific class $c \in K$. Then, the video classifier $\mathcal{C}$ maps an adversarial video $\bar{Y}$ to the class with the maximum probability. Thus, the adversarial loss $\mathcal{L}_{class}$ for the untargeted and targeted attacks can be obtained from

\vspace{-2mm}
\begin{equation} \label{eq:eq2}
\min_{\Delta} \ \mathcal{L}_{class}(X, \Delta, \lambda), \quad\quad
\mathcal{L}_{class}(X, \Delta, \lambda) =\begin{cases}
  F_{\mathcal{C}(Y)}(\bar{Y}) - \max\limits_{c \neq \mathcal{C}(Y)}F_{c}(\bar{Y}) & \text{(Untargeted)}\\
  \max\limits_{c \neq c^{*}}F_{c}(\bar{Y}) - F_{c^{*}}(\bar{Y}) & \text{(Targeted)}
\end{cases}
\vspace{-1.5mm}
\end{equation}

\textbf{Undetectability Constraint.} We incorporate two regularization terms ($\mathcal{R}_{thick}, \mathcal{R}_{rough}$) proposed by \cite{pony2021over} to craft the imperceptible flickering perturbations, where $\mathcal{R}_{thick}$ denotes the magnitude of perturbations, and $\mathcal{R}_{rough}$ is the amount of change in flickering perturbations between adjacent frames. We obtain each regularization term as follows:
\vspace{-1mm}
\begin{equation} \label{eq:eq3}
\mathcal{R}_{thick}(\Delta) = \ \frac{1}{3T} \norm{\Delta}^{2}_{2}, \quad\quad
\mathcal{R}_{rough}(\Delta) = \ \frac{1}{3T} (\norm{\pd{\Delta}{t}}^{2}_{2} + \norm{\pdd {\Delta} t}^{2}_{2})
\vspace{-1.5mm}
\end{equation}

where $\norm{(\cdot)}_{2}$ is a Tensor p-norm defined in \cite{pony2021over}. $\pd {\Delta} {t}$ can be obtained by $\Gamma(\Delta, 1) - \Gamma(\Delta, 0)$, where a permutation function $\Gamma(\Delta, \tau)$ produces a cyclic temporal shift of the original perturbation $\Delta_{g}$ by an offset $\tau$. We find the second order temporal derivative from $\pdd {\Delta} t = \Gamma(\Delta, -1) - 2\Gamma(\Delta, 0) + \Gamma(\Delta, 1)$ .

\textbf{Objective Function.} In the offline attack scenario, injecting the adversarial
perturbations is not latency bound. The adversary can therefore formulate the below adversarial function to minimize the adversarial loss:
\vspace{-1mm}
\begin{equation} \label{eq:eq4}
\resizebox{0.93\columnwidth}{!}{
$\begin{aligned}
\min_{\Delta}  \ \sum^{\floor{T/G}}_{g=0} \frac{\mathcal{L}_{comp}(X, \Delta_{g}, \lambda, g)}{\floor{T/G}+1} + \beta \mathcal{L}_{class}(X, \Delta, \lambda) + \zeta(\mathcal{R}_{thick}(\Delta) + \mathcal{R}_{rough}(\Delta)) \quad
\text{s.t.,} & \norm{\Delta}_{\infty} \leq \epsilon,
\end{aligned}$}
\vspace{-1.5mm}
\end{equation}

where $\beta$ adjusts the scale of the two loss functions. $\zeta$ determines the importance of $\mathcal{R}_{thick}$ and $\mathcal{R}_{rough}$. To ensure the injected noise is imperceptible to humans, the norm of the perturbation is upper bounded by a pre-defined small value $\epsilon$.

\subsection{Online Attacks}
\label{sec:onlineattack}
For online attack settings, we assume that the adversary does not have access to any user data, meaning that a per-video perturbation cannot be formed. Therefore, the adversary may not be able to align the flickering perturbations with the desired video sequences. To address this challenge, we follow RoVISQ~\cite{chang2022rovisq} to 
craft universal perturbation that has input-invariant characteristics using the permutation function $\Gamma(\Delta, \tau)$. We also set the temporal length of the perturbation to the GOP size ($G$). 
Considering a black-box attack scenario,
we demonstrate the extent to which our universal perturbations trained on a surrogate open-source video compression model, are transferable to unseen models and architectures. These universal perturbations are trained to minimize the adversarial loss in Equation~\ref{eq:eq4} using a publicly available training dataset.


\section{Attack Evaluation} \label{sec:eval}

\subsection{Evaluation Setup}
\textbf{Victim Model} We evaluate our attacks on the state-of-the-art video compression model DVC~\cite{Lu_2019_CVPR}. In a video compression pipeline, the first video frame is always encoded using DNN-based image compression~\cite{liu2020unified}. Video quality is measured as peak signal-to-noise ratio (PSNR) and the bit-rate is calculated by bits per pixel (Bpp). For downstream video classification, we further extend our attack to three state-of-the-art video classification models, specifically I3D~\cite{carreira2017quo}, SlowFast~\cite{feichtenhofer2019slowfast}, and TPN~\cite{Yang_2020_CVPR}.

\textbf{Dataset} We use the Vimeo-90K dataset for training the victim video compression model. We set the GOP size to 10 following prior work~\cite{Lu_2019_CVPR}. \sys is evaluated on the hand gesture recognition dataset 20BN-JESTER (Jester)~\cite{materzynska2019jester}. We split the Jester dataset into train, validation, and test set in the ratio 8:1:1. We set our adversarial perturbation bound to $\epsilon=0.2$ by following prior attack proposed by ~\cite{pony2021over}.

\subsection{Experimental Results}
\begin{figure}[t!]
\centering
  \begin{tabular}{@{}c@{}}
    \includegraphics[width=0.6\linewidth]{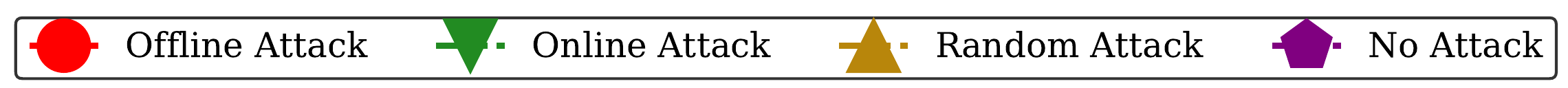} \\  
  \end{tabular}
    \begin{tabular}{@{}cccc@{}}
    \includegraphics[width=0.24\linewidth]{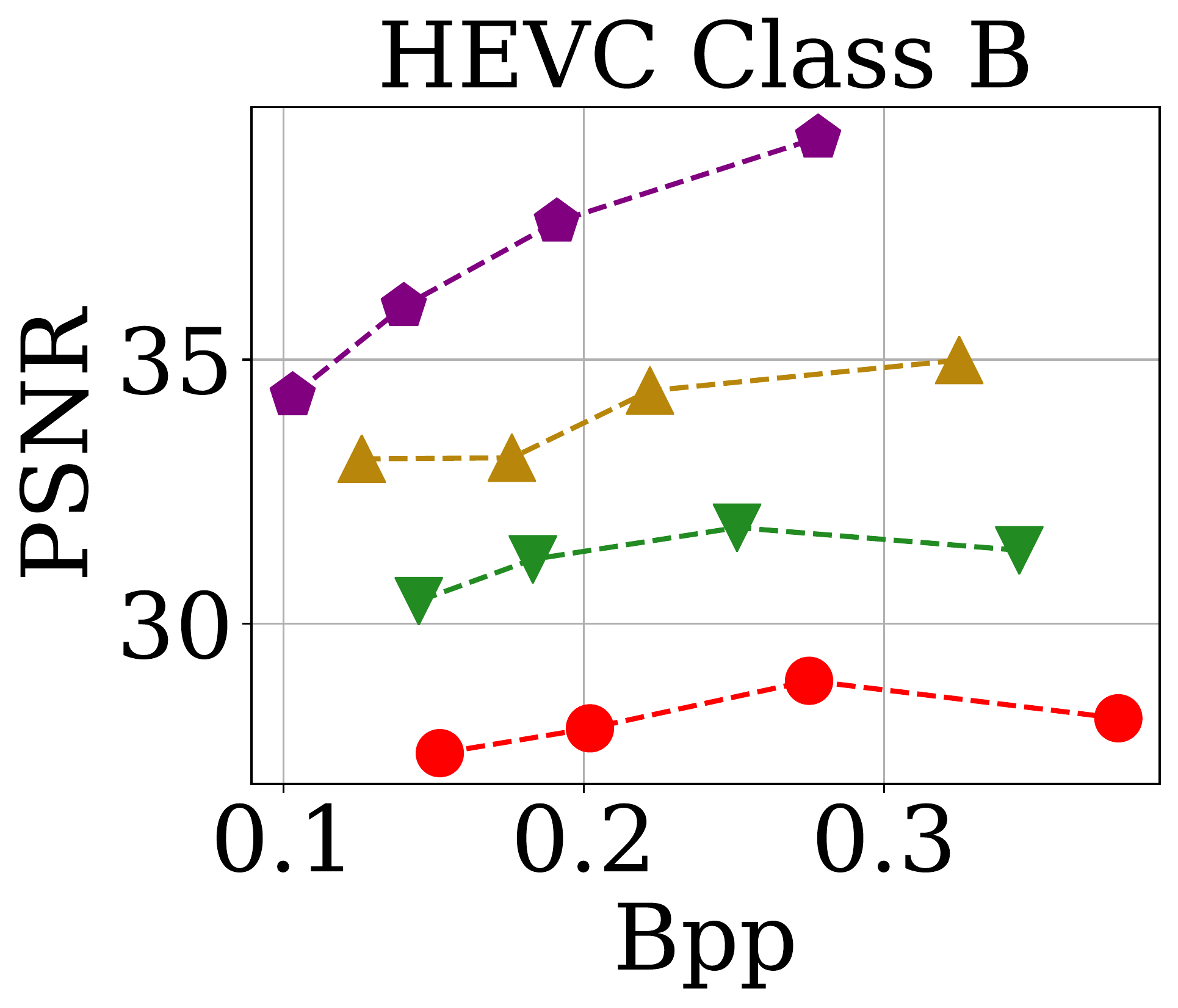}   &
    \includegraphics[width=0.24\linewidth]{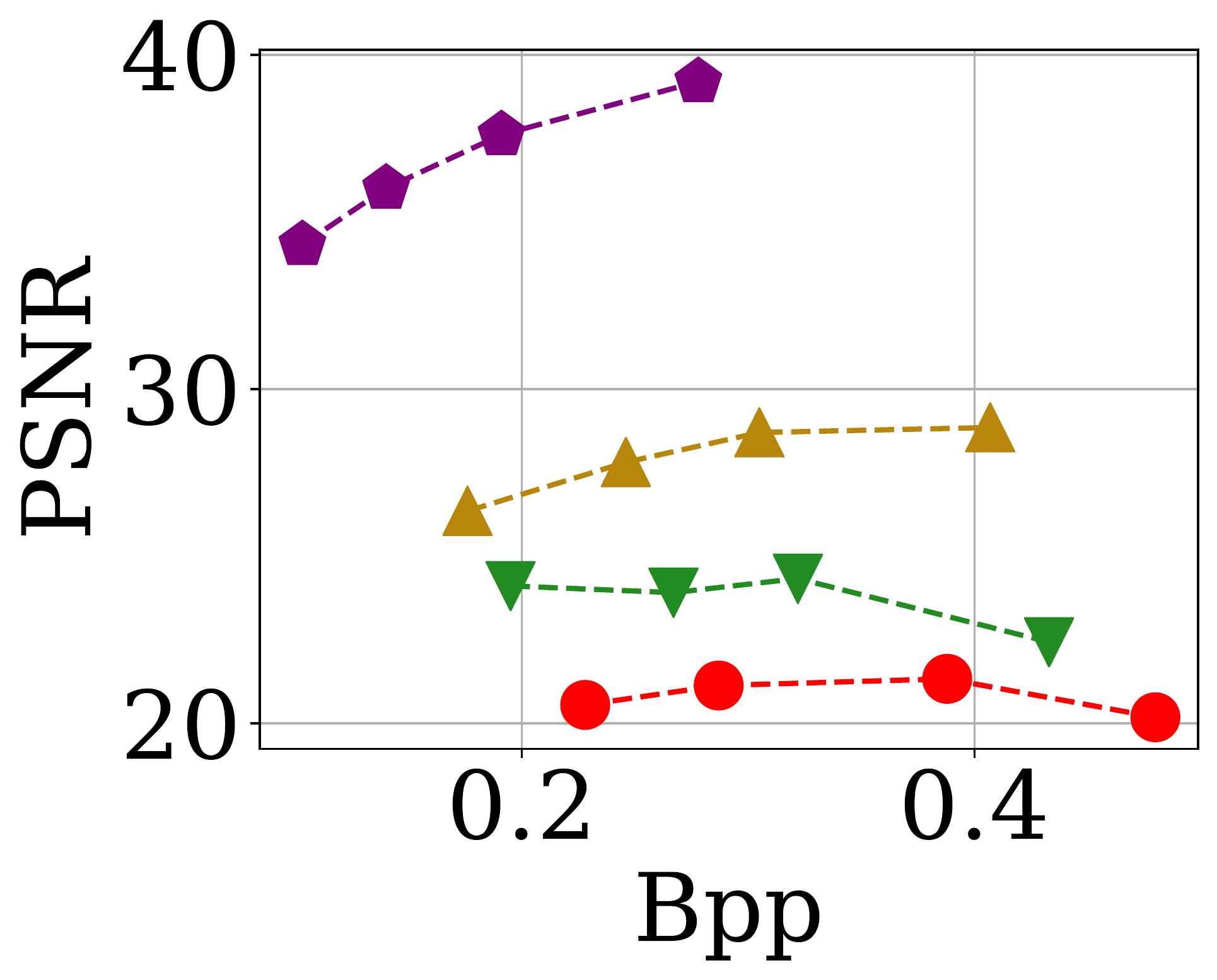}   &
    \includegraphics[width=0.24\linewidth]{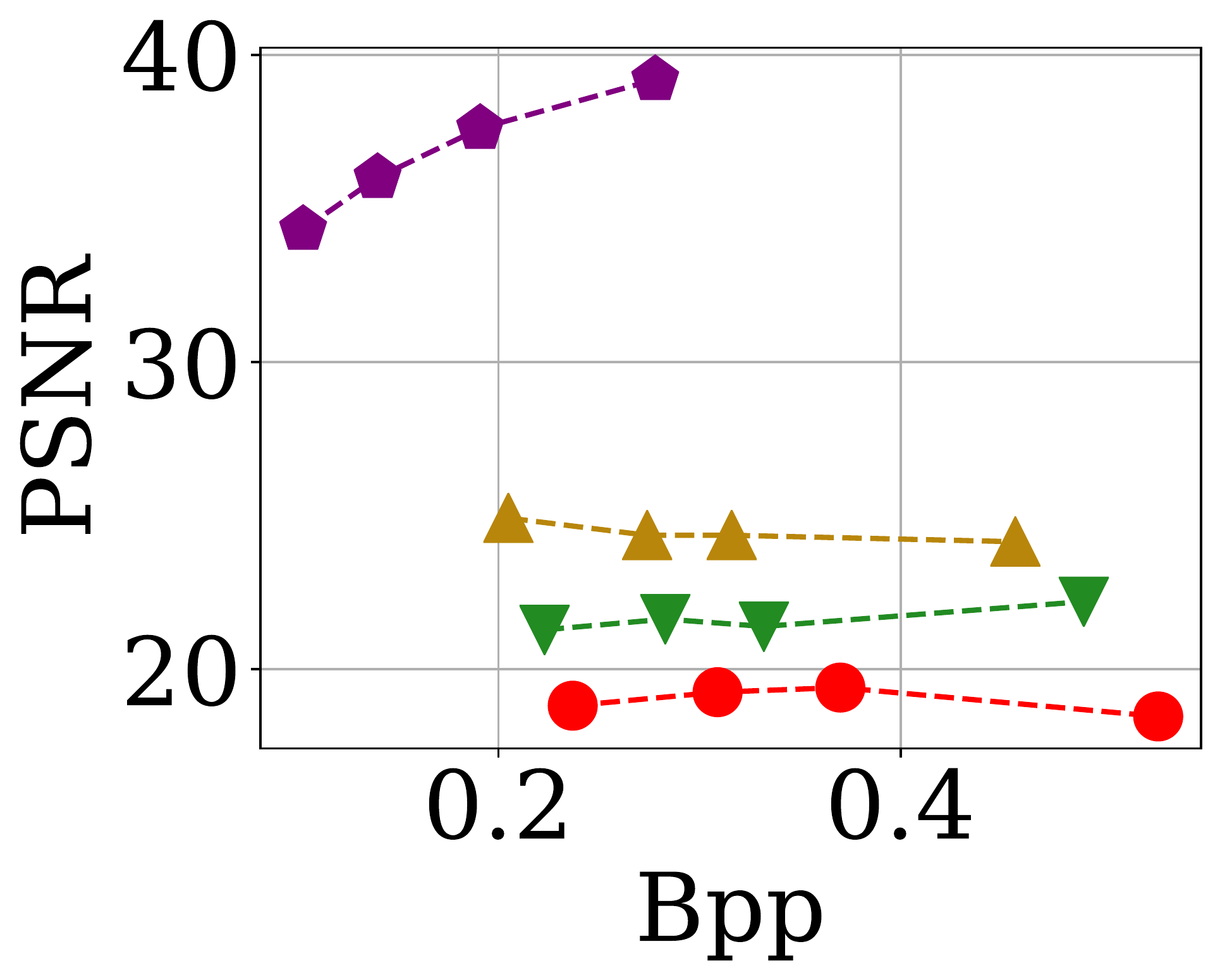}   &
    \includegraphics[width=0.24\linewidth]{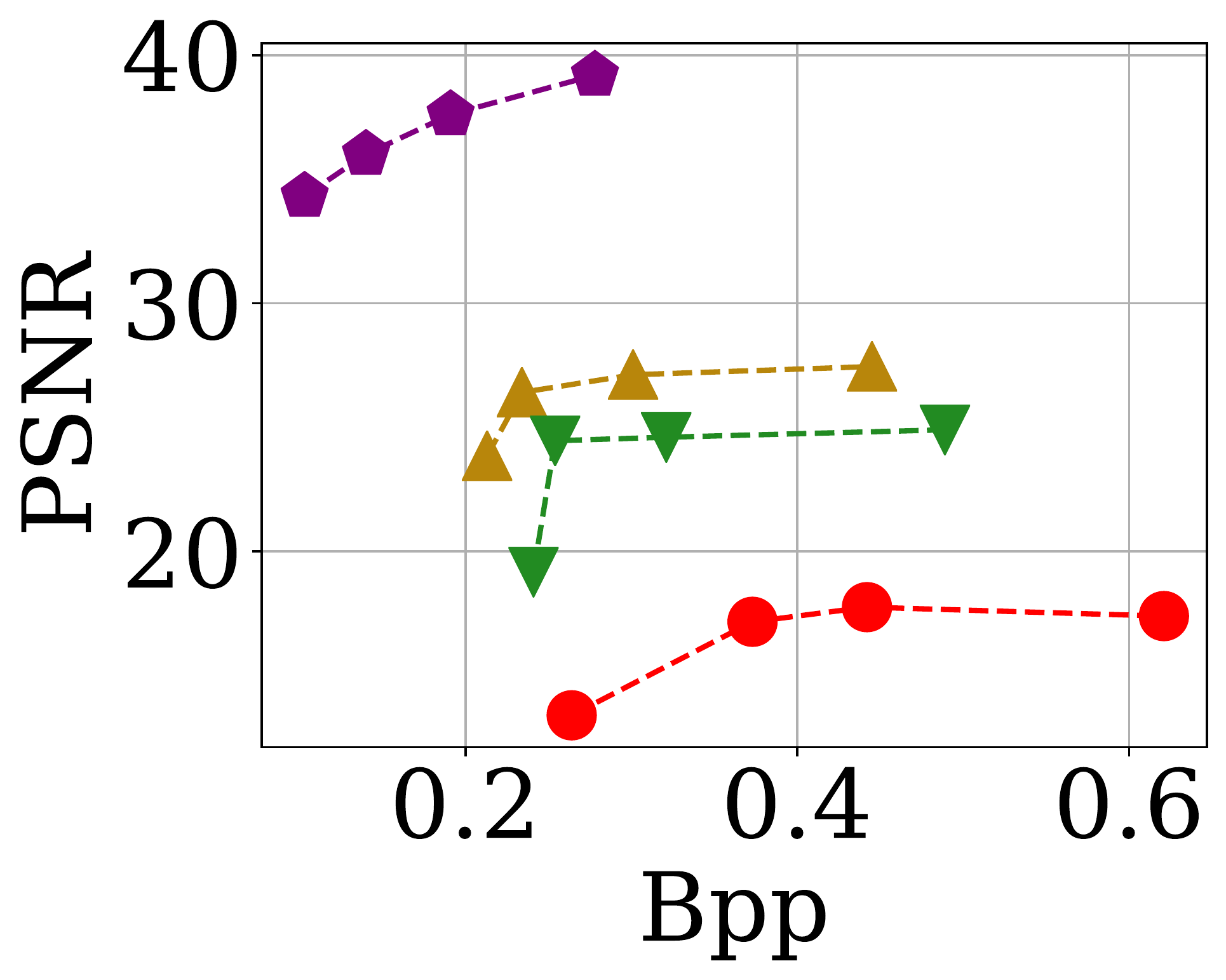} \\
    \footnotesize{(a) $\epsilon=0.04$}&
    \footnotesize{(b) $\epsilon=0.12$}&
    \footnotesize{(c) $\epsilon=0.16$}&
    \footnotesize{(d) $\epsilon=0.20$} \\
    \end{tabular}  
  \caption{\sys applied to DVC~\cite{Lu_2019_CVPR} video compression framework with different values of $\epsilon$. Each graph contains the results with different $\lambda \in \{256,512,1024,2048\}$.}
  \label{fig:vc_result}
  \vspace{-0.5cm}
\end{figure}

\textbf{Video Compression.} Figure~\ref{fig:vc_result} shows the PSNR-based \textit{R-D} performance evaluated on the Jester dataset. We change the values of $\epsilon$ to analyze how the attack performance is affected by the $l_{\infty}$ norm of the flickering perturbations. Each point in the Figure~\ref{fig:vc_result} represents the result (PSNR, Bpp) of video compression according to a total of four $\lambda$ values ($\lambda \in \{256,512,1024,2048\}$). We observe that the attack causes a larger drop in both PSNR and Bpp for smaller values of $\lambda$. This is because the compression rate decreases as $\lambda$ increases, so that the video compression model encodes less information in the current frame. Furthermore, the attack is more effective because there is less loss to the perturbations present in the frame. Specifically, our attack increases Bpp by up to 2.23$\times$ and results in distortion up to 25.81dB. In addition, we compare our attacks with uniformly sampled noise $\Delta_{U} \sim \mathcal{U}(-\epsilon,\epsilon)$. Injecting random noise to videos can only achieve -10.47dB and 1.6$\times$ on average. When applying our online perturbations to unseen models (black-box attack), the average PSNR drop and Bpp increase are -13.01dB and 2.34$\times$ on average.

\begin{table}[t!]
\centering
\caption{Accuracy (ACC) of video classification on clean videos, along with the attack success rate (ASR) of flickering perturbations. The defense is adversarial training on the Jester dataset. Here, ``T'' and ``U'' denote targeted and untargeted attacks, respectively. } \label{tab:vr_result}
\resizebox{\columnwidth}{!}{
\begin{tabular}[t]{ccccccc|c}
\toprule
\begin{tabular}[c]{@{}c@{}}Video\\ Classifier\end{tabular}  
& Type
& Dataset
& $\epsilon$
& Attack
& Surrogate
& ASR (\%) 
& ACC (\%)
\\
\midrule
\multirow{3}{*}{\begin{tabular}[c]{@{}c@{}}SlowFast\\ \cite{feichtenhofer2019slowfast}\end{tabular}}
& T & \multirow{3}{*}{Jester} & \multirow{3}{*}{0.2} & Offline & - & 92.6 & \multirow{3}{*}{89.5}\\
& U & & & Offline & - & 96.3 & \\
& U & & & Online & TPN & 83.3 & \\ \hline
\multirow{3}{*}{\begin{tabular}[c]{@{}c@{}}TPN\\ \cite{Yang_2020_CVPR}\end{tabular}}
& T & \multirow{3}{*}{Jester} & \multirow{3}{*}{0.2} & Offline & - & 93.5 & \multirow{3}{*}{90.5} \\ 
& U & & & Offline & - & 97.2 & \\
& U & & & Online & I3D & 86.1 & \\ \hline
\multirow{3}{*}{\begin{tabular}[c]{@{}c@{}}I3D\\ \cite{carreira2017quo}\end{tabular}}
& T & \multirow{3}{*}{Jester} & \multirow{3}{*}{0.2} & Offline & - & 95.3 & \multirow{3}{*}{91.2} \\ 
& U & & & Offline & - & 98.1 & \\
& U & & & Online & SlowFast & 85.1 & \\
\bottomrule
\end{tabular}}
\vspace{-0.5cm}
\end{table}

\begin{figure}[t!]
\centering
  \begin{tabular}{@{}c@{}}
    \includegraphics[width=0.6\linewidth]{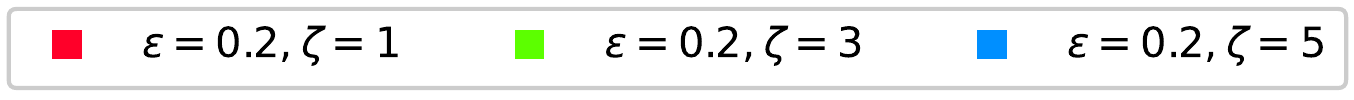} \\  
  \end{tabular}
    \begin{tabular}{@{}ccc@{}}
    \includegraphics[width=0.32\linewidth]{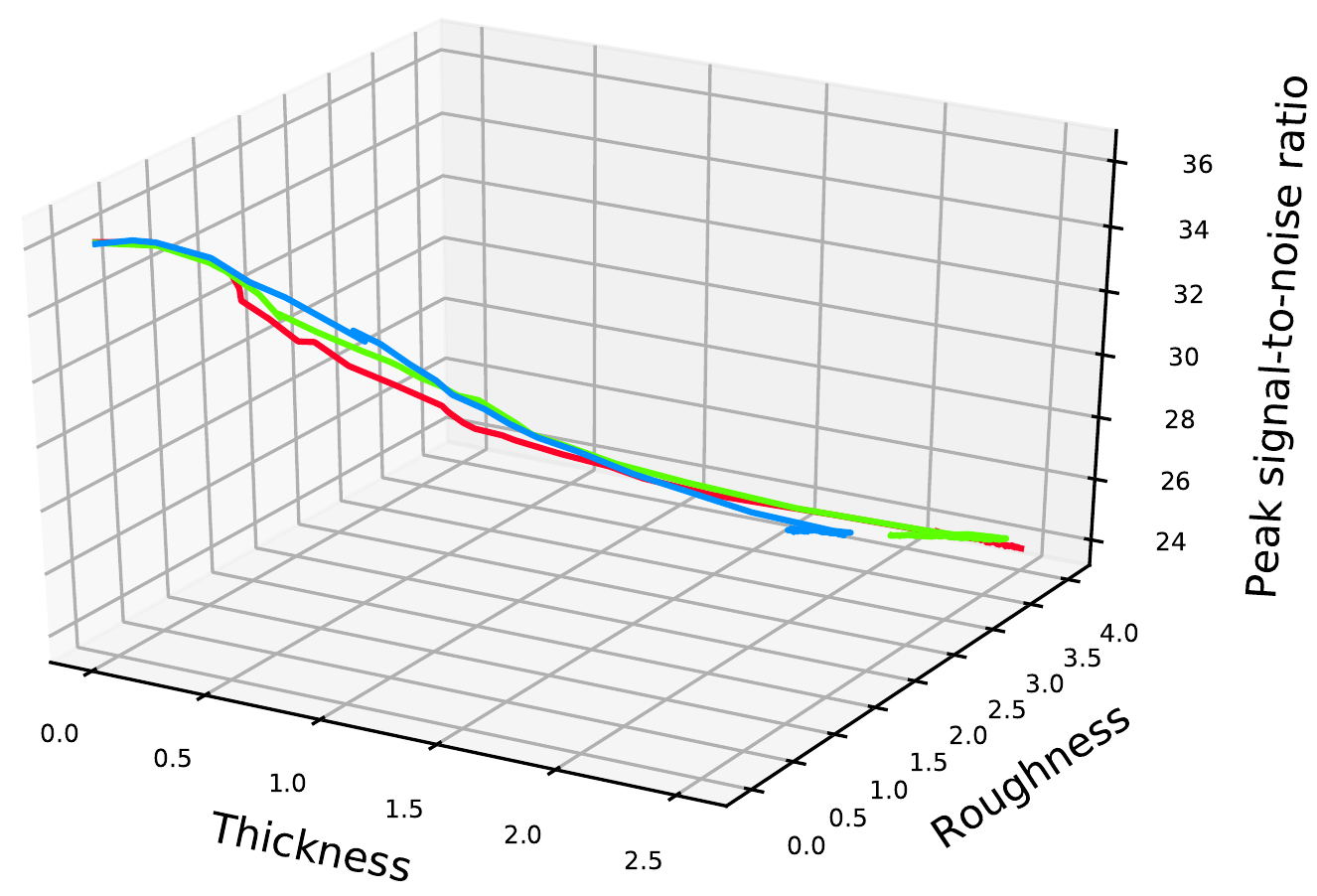}   &
    \includegraphics[width=0.32\linewidth]{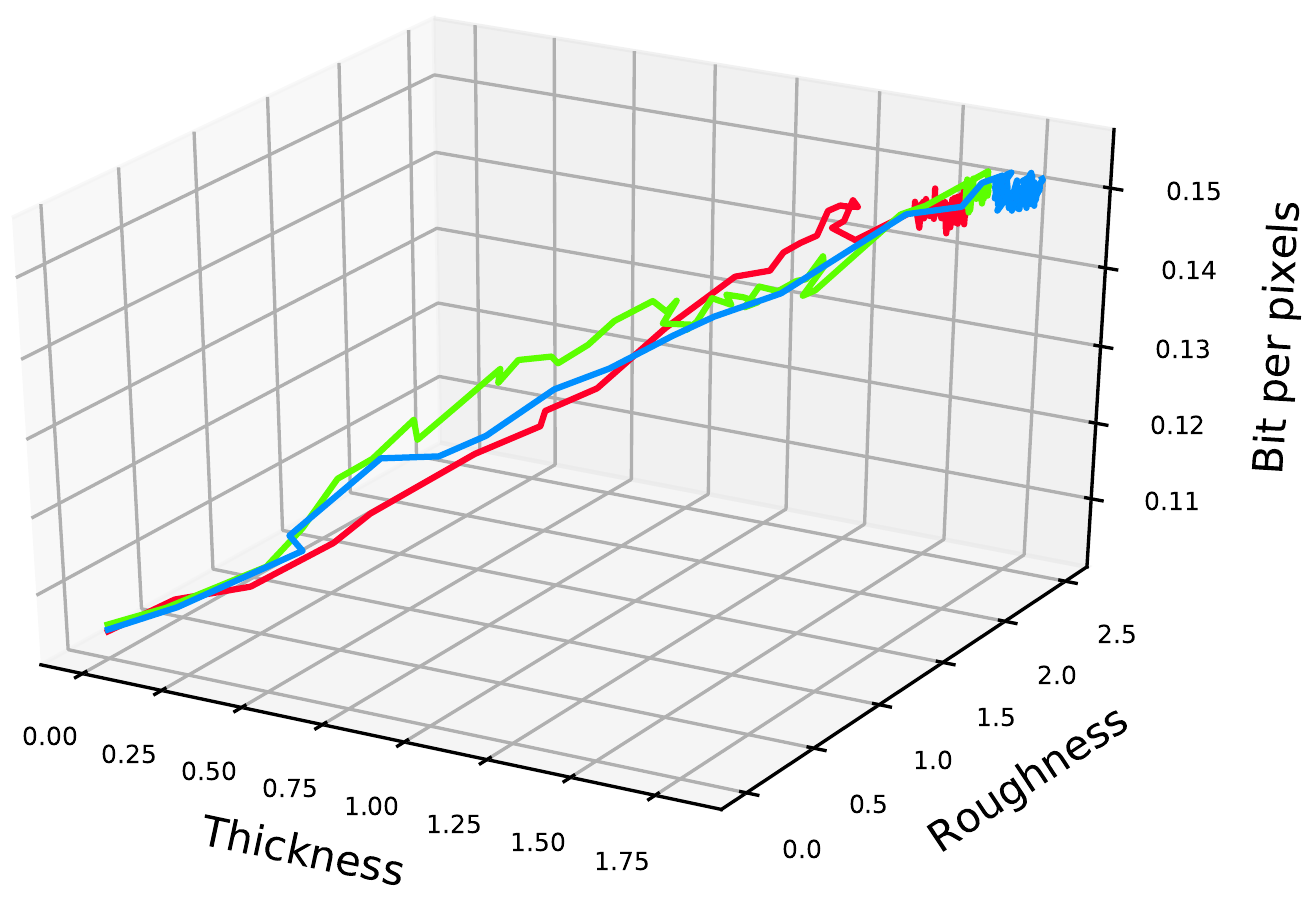}    &
    \includegraphics[width=0.32\linewidth]{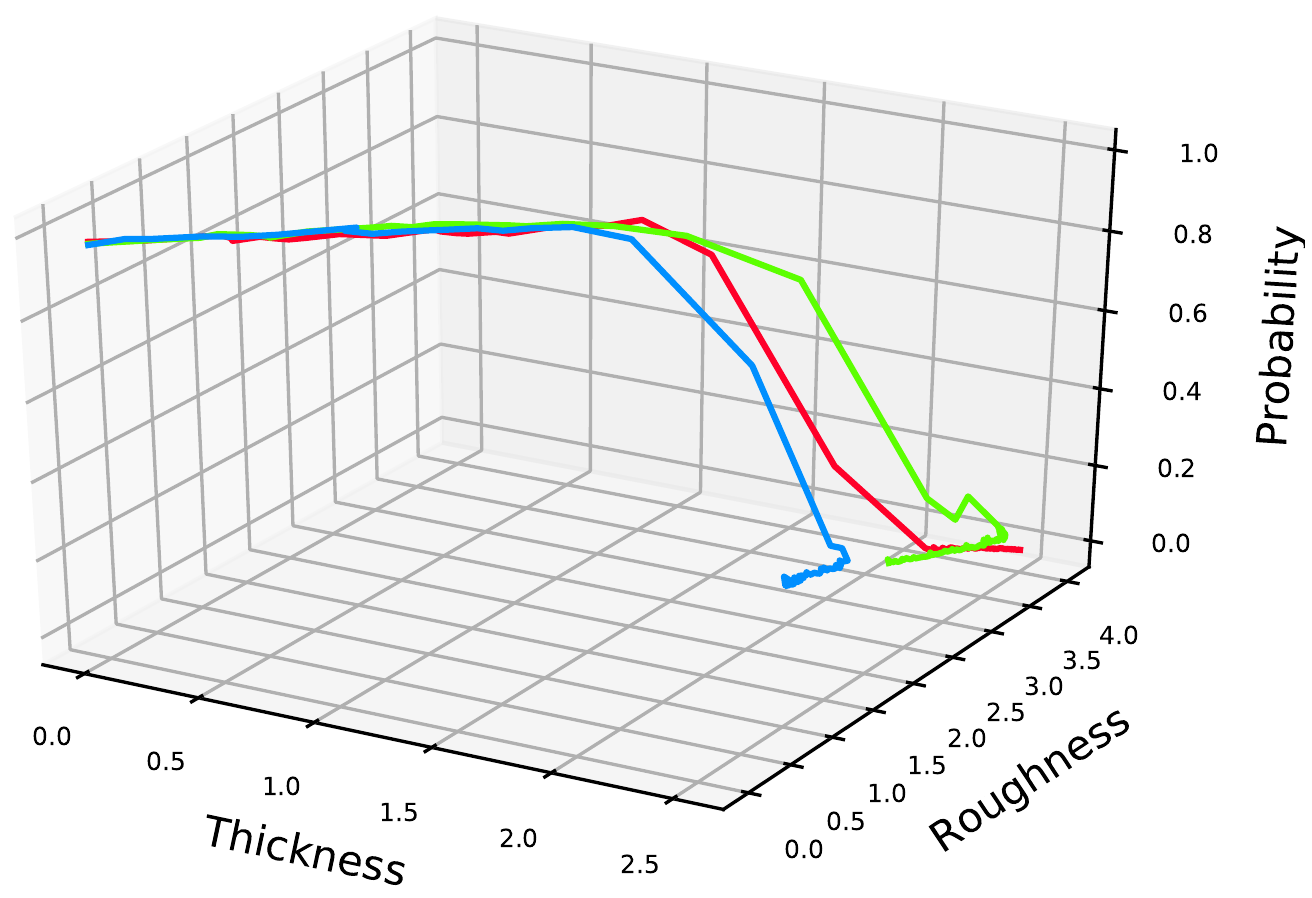} \\
    \footnotesize{(a)}&
    \footnotesize{(b)}&
    \footnotesize{(c)} \\
    \end{tabular}  
  \caption{Convergence curve in the search space of \sys. (a) PSNR-thickness-roughness, (b) Bpp-thickness-roughness, (c) Probability-thickness-roughness.}
  \label{fig:convergence}
  \vspace{-0.3cm}
\end{figure}

\begin{figure}[t!]
\centering
    \begin{tabular}{@{}cc@{}}
    \includegraphics[width=0.54\linewidth]{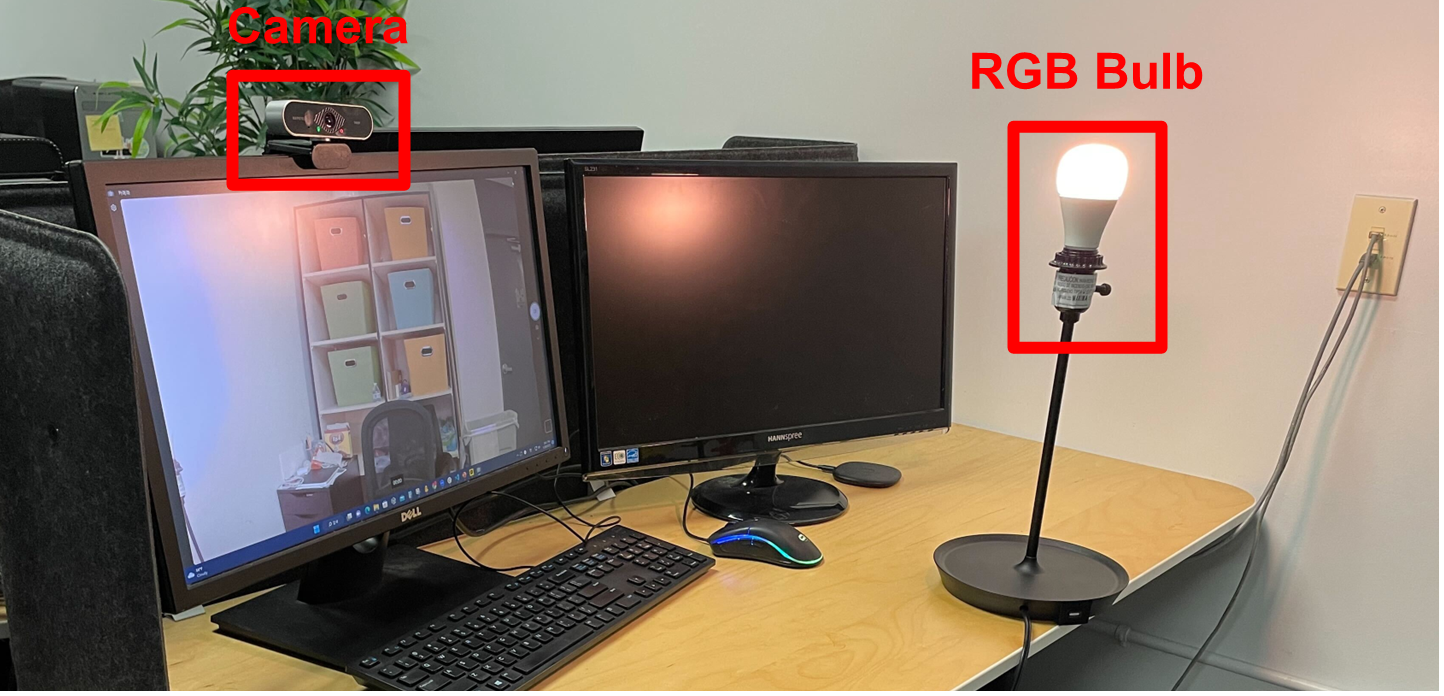}   &
    \includegraphics[width=0.41\linewidth]{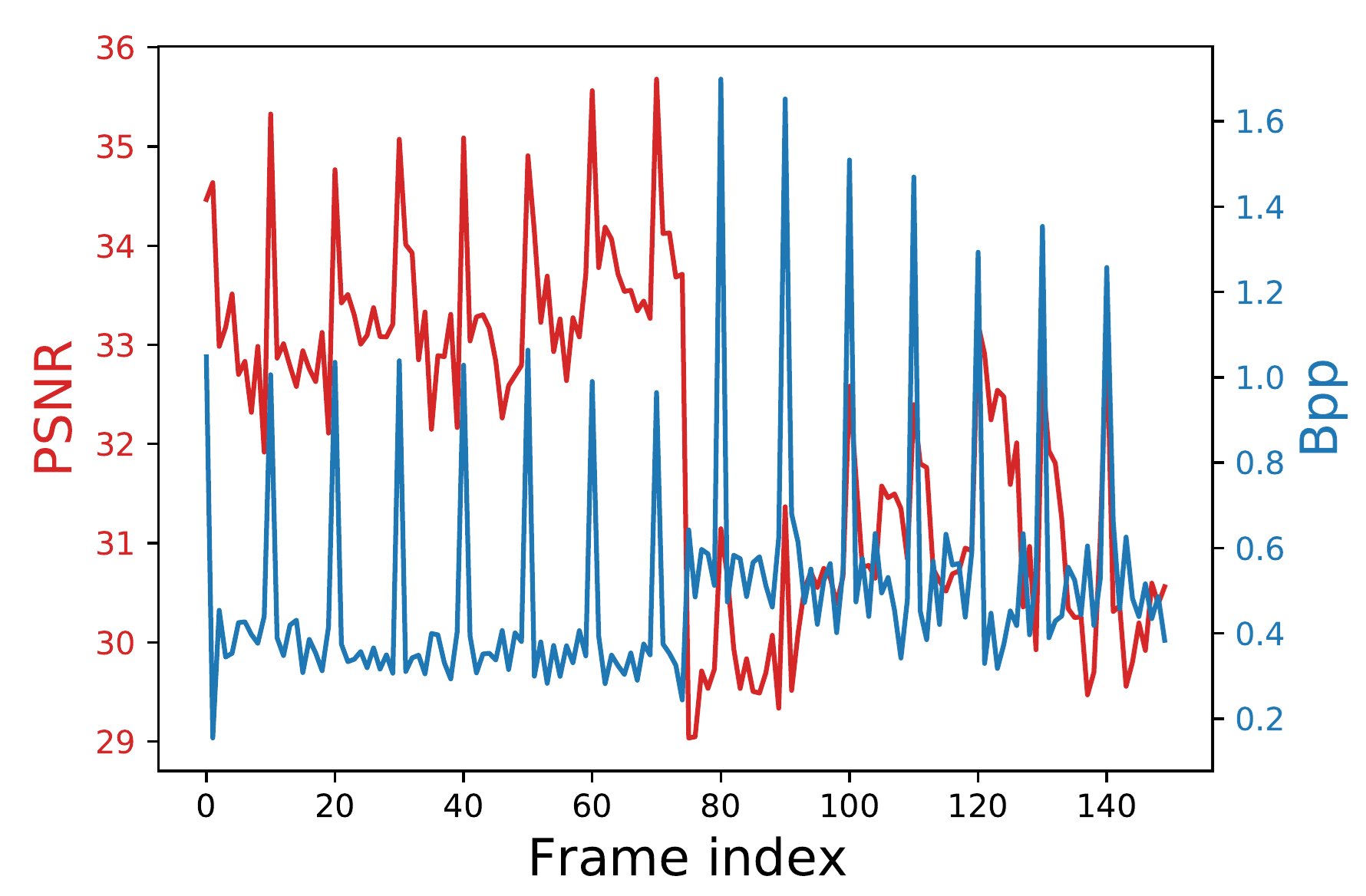} \\
    (a) &
    (b) \\
    \end{tabular}  
    \vspace{-0.2cm}
  \caption{(a) Experimental setups for evaluating \sys in the physical domain, (b) Visualization of PSNR and Bpp of continuous video frames in a situation where the attack starts from the 75th frame.}
  \label{fig:demo}
  \vspace{-0.5cm}
\end{figure}

\textbf{Downstream Video Classification.} We evaluate the performance of offline and online attacks respectively, when the back-end user queries the video classifiers for analysis. We measure the attack success rate (ASR) by counting misclassified test videos for the untargeted attack. For the targeted attack, the success rate is the portion of the sample mapped to the class the adversary wants. We summarizes the ASR of \sys on Jester dataset in Table~\ref{tab:vr_result}. In the offline attack settings, we obtain over 90$\%$ ASR. We also consider an online attack scenario, where the video sequences continuously captured by a camera are classified in real-time. Even in such real-time scenarios, our online attacks that inject universal perturbations into the video are highly effective with 86.1$\%$ ASR.

\textbf{Convergence Process.} Figure~\ref{fig:convergence} shows how our flickering perturbations converge to the optimization point considering several terms. We can see that the adversarial perturbation drastically shifts the starting point(Initial BPP, PSNR and Probability) of the optimization function to a path that significantly lowers the R-D performance and then stops to reflect the regularization terminals.

\subsection{Real-world Deployment} We test \sys in the real-world using a Kasa KL130 Smart Bulb~\cite{Kasa}, which supports a wide array of colors and can be controlled by Wi-Fi. As shown in Figure~\ref{fig:demo} (a), we conduct an experimental design to physically test the effect of RGB bulb on video compression. Figure~\ref{fig:demo} (b) shows the visualization of the PSNR and Bpp of continuous video frames, where online attacks are performed by attackers from the 75th frame. Note that video compression encodes the first frame of each GOP to a higher Bpp than the others for the efficiency of spatio-temporal prediction. Comparing between the same frame types (e.g., I-frame, P-frame), we observe that \sys successfully degrades video quality and compression rates by up to 6.5dB and 1.6$\times$, respectively.

\vspace{-1.5mm}
\section{Conclusion and Future Works}

We present \sys, the first physical adversarial attack to DNN-based video compression pipelines in the real-world. \sys utilizes a simple, yet effective, setup to project adversarially crafted flickering perturbations onto a victim IoT camera. We show \sys's efficacy by evaluating the attack's effects on both PSNR and Bpp when applied to a state-of-the-art video compression framework. Our results indicate that such an attack not only causes a significant drop in video compression performance, but also achieves high attack success rate on downstream video classifiers. Finally, we show the real-world applicability of our physical attack on video compression, by utilizing an RGB LED light bulb to project imperceptible adversarial patterns in real-time onto camera sensors capturing video. This work presents an important first step in a paradigm shift towards physical and realizable adversarial attacks on IoT video streaming devices. In future work, we aim to build a strong defense against \sys that will further secure the video compression and classification pipeline in IoT, ensuring security and robustness against attacks.



\section*{Acknowledgement}
\vspace{-1.5mm}
This work was supported by the U.S. Army/Department of Defense under award number W911NF2020267 and ARO MURI under award number W911NF-21-1-0322.

\bibliography{iclr2023_conference}
\bibliographystyle{iclr2023_conference}


\end{document}